\documentclass{aa}
\usepackage{graphicx}
\usepackage{lscape}
\usepackage{longtable}
\usepackage{txfonts}
\usepackage{natbib}

\begin{document}

   \title{PHOENIX model chromospheres of mid- to late-type M \mbox{dwarfs \thanks{Based on observations
collected at the European Southern Observatory, Paranal, Chile, 68.D-0166A.},\,\thanks{Table
A.2 is only available in electronic form at the CDS via anonymous ftp.}} }

   \titlerunning{PHOENIX model chromospheres}

   \author{B. Fuhrmeister,
          J. H. M. M. Schmitt \and P. H. Hauschildt
          }

   \offprints{B. Fuhrmeister}

   \institute{Hamburger Sternwarte, University of Hamburg,
              Gojenbergsweg 112, D-21029 Hamburg\\
              \email{bfuhrmeister@hs.uni-hamburg.de}
             }


   \abstract{We present semi-empirical model chromospheres computed with 
             the atmosphere code PHOENIX. The models are designed to fit
             the observed spectra of five mid- to late-type M dwarfs. Next
             to hydrogen lines from the Balmer series we used various metal
             lines, e.\,g. from Fe\,{\sc i}, for the comparison between data and models. 
             Our computations show that an NLTE treatment of C,\,N,\,O 
             impacts on the hydrogen line formation, while NLTE treatment of
             less abundant metals such as nickel influences  the lines
             of the considered species itself. For our coolest models we
             investigated also the influence of dust on the chromospheres and
             found that dust increases the emission line flux. 

             Moreover we present an (electronically published) emission line list for
             the spectral range of 3100 to 3900 and 4700 to 6800 \AA\, for a
             set of 21 M dwarfs and brown dwarfs. The line list includes the detection of the 
             Na\,{\sc i} D lines in emission for a L3 dwarf.

   }

   \maketitle
%

\section{Introduction}

Chromospheric activity as indicated, for example, by H$_{\alpha}$ emission is frequently found in early-type
M dwarfs, and ubiquitously in mid to late-type M dwarfs. There are indications that the H$_{\alpha}$ emission
 during quiescence declines for very late-type M dwarfs and L dwarfs \citep{Liebert1, Gizis},
 although even brown dwarfs
can show H$_{\alpha}$ emission  at least during flares. Since the heating mechanisms of chromospheres and coronae are poorly
understood, we may hope to learn more about the observed emission lines via the construction of semi-empirical
chromospheres. 

Semi-empirical modelling of the chromosphere of the Sun
was carried out quite successfully by \citet{Vernazza}, determining the temperature distribution versus
the column mass ($m$). Early models for M dwarfs
were constructed by \citet{Cram}. More recently, \citet{Hawley} constructed chromosphere
and transition region (TR) models of different activity level including soft X-ray emission
from the corona. An ansatz using a linear temperature rise vs. log $m$ in the chromosphere
and TR was used
by \citet{Short} and related papers (e.\,g. \citet{Short1}, \citet{Andretta}).
These investigators used the atmospheric code MULTI \citep{Carlsson}, and
in addition the atmospheric code PHOENIX \citep{phoenix} 
to calculate background opacities. \citet{Falchi} instead used the atmosphere
code Pandora \citep{Pandora} and a non-linear temperature versus log $m$ distribution.
The lines under consideration were usually the Ly$_{\alpha}$ line, the H$_{\alpha}$ line,
the Ca\,{\sc ii} H and K lines and a few other metal lines. One problem with such models lies
in the uniqueness of the description. In other words: Can there be two different models
producing the same line fluxes? Naturally this problem must be more severe if only few lines
are used for adjusting the model. 

With the advent of large telescopes like the VLT it is now
possible to obtain high quality spectra in the (optical) near UV, where M dwarfs exhibit hundreds of
chromospheric emission lines. Since most of these lines are Fe\,{\sc i} and {\sc ii} lines, at least
Fe has to be computed in NLTE in addition to H and He -- an approach that has become possible in the last few years
due to  increasing computing power.

In this paper we present model chromospheres for mid-type M dwarfs during quiescent state
adjusted via various lines in the wavelength range between 3600 and 6600 \AA. 
Our paper is structured as follows:
In sect. 2 we describe the VLT data used for our analysis and the sample of M dwarfs.
In sect. 3 we deal with the model construction and describe the influence of various
model parameters in sect. 4. We present our best fit models for the
individual stars in sect. 5 and discuss several aspects of the models in sect. 6.
In the appendix a catalog of chromospheric emission lines is presented.


\section{Observations and data analysis}

A set of 23 M dwarf spectra was taken with UVES/VLT in visitor and in service mode
between winter 2000 and March 2002. The original sample was designed for a search for
the forbidden Fe\,{\sc xiii} line at 3388 \AA\, and covers the whole M dwarf regime from M 3.5
plus a few L dwarfs known to show H$_{\alpha}$ activity. All stars were selected for their
high activity level. Two of the stars were double stars and since the spectra could not be
disentangled, they were excluded  from analysis, and we ended up with 21 objects. 

The
five objects used in the modelling were selected  
to cover the M dwarf temperature regime
with good S/N ratios  and without obvious
flaring activity during the observations.  
The spectra
 were obtained in visitor mode with 
ESO's Kueyen telescope at Paranal equipped with the Ultraviolet-Visual Echelle 
Spectrograph (UVES) from March, 13th to 16th in 2002. The instrument 
was operated in  dichroic mode, yielding
33 echelle orders in the blue arm (spectral coverage from 3030 to 3880 \AA) and 39 orders
in the red arm (spectral coverage from 4580 to 6680 \AA). 
Therefore
we cannot observe the lines from H$_{3}$ up to H$_{8}$ of the Balmer series,
nor do we cover the Ca\,{\sc ii} H and K lines.  
The typical resolution of our spectra is $\sim 45000$, typical exposures lasted 5 to 20 minutes. 
Unfortunately the H$_{\alpha}$ line is saturated in all of our spectra for AD~Leo, CN~Leo
and YZ~CMi. On the other hand, the blue part of the spectrum is underexposed for LHS~3003
and partly for DX~Cnc, and therefore was not used for the modelling.

All data were reduced using IRAF in a standard way. 
The wavelength calibration was carried out using Thorium-Argon spectra with
an accuracy of $\sim 0.03$\AA\,in the blue arm and $\sim 0.05$\AA\,in the red arm (i.\,e.,
more than 90 percent of the residuals of the wavelength calibration are lower than this value; 
the same
is found for the difference between measured and laboratory wavelength in the 
emission line measurements presented in the appendix).
In addition to the UVES spectra there are photometric data from the UVES exposure meter.  
These data were actually taken for engineering purposes, and are therefore not flux calibrated. 
Still, these data were useful to assess 
whether the star was observed during quiescence or during a major flare. We used the photometer
data to decide which spectrum was taken during quiescence and therefore can be used 
for the chromospheric modelling. If no flare occurred during the whole observation and the spectra 
seem to be quite stable we used the averaged spectrum to obtain a better S/N (see \mbox{Tab. \ref{phot}}).

\begin{table*}
\caption{\label{phot}Parameters of the underlying photospheres for the model construction. From
the literature we also cite [A/H] given in brackets.}
\begin{tabular}[htbp]{cccccc}
\hline
star & other name & spectrum used & T$_{\mathrm{eff}}$ [K] & log g & literature\\
\hline
AD Leo & GJ 388 & 2002-03-16 03:40:47 & 3200 & 4.5 & 3350, 4.5 (-0.75) $^{1}$\\
YZ CMi & GJ 285 & average & 3000 & 4.5 & 2925 $^{3}$\\
CN Leo & GJ 406 & 2002-03-14 03:24:38 & 2900 & 5.5 & 2900, 5.0 (0.0) $^{1}$\\
DX Cnc & GJ 1111 & average &2700 & 5.0 & 2850, 5.25 (+0.5) $^{1}$; 2775 $^{3}$\\
LHS 3003 & GJ 3877 & average &2500 & 4.5 & 2400-2650 $^{2}$\\
\hline
\end{tabular}
\\
$^{1}$ \citet{Jones} \hspace*{1cm}
$^{2}$ \citet{Leggett} \hspace*{1cm}
$^{3}$ \citet{Mohanty}
\end{table*}

\begin{figure} 
\begin{center}
\includegraphics[width=8cm]{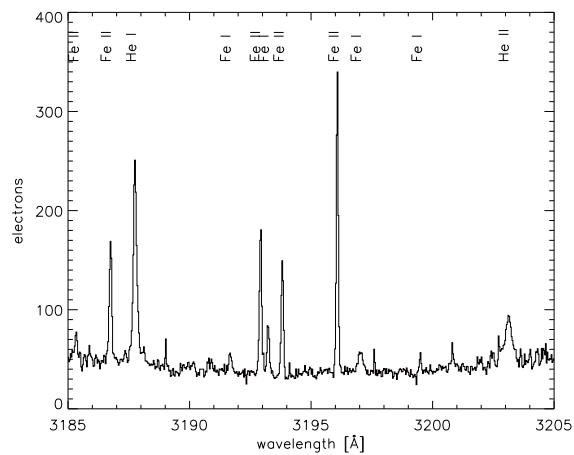}
\caption{\label{lines1}A typical blue part of the spectrum of CN~Leo around 3200 \AA. 
The emission lines
belong to Fe\,{\sc i}, Fe\,{\sc ii}, He\,{\sc i} and He\,{\sc ii}.}
\end{center}
\end{figure}

\begin{figure} 
\begin{center}
\includegraphics[width=8cm]{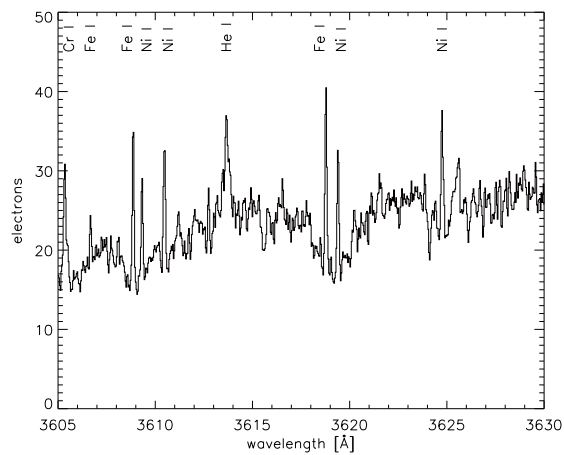}
\caption{\label{lines}Same as is Fig. \ref{lines1} around 3600 \AA.  The emission lines
belong to Ni\,{\sc i}, Fe\,{\sc i}, Cr\,{\sc i} and He\,{\sc i}.}
\end{center}
\end{figure}

We show two typical parts of the spectrum of CN~Leo in Figs. \ref{lines1} and \ref{lines}
to point out the wealth of emission lines in the blue part of the
spectrum. 
There are hundreds of emission lines originating in the chromosphere, all of which could 
in principle be used for the modelling. 
 Since the lines belong to many different species we had to decide which
species to use since not all of them could be calculated in NLTE if computation times are to
remain reasonable (see section \ref{construction}). Identifications of the emission lines
show that by far most of them are Fe\,{\sc i} lines. For more detailed informations on the
emission lines seen on the different stars we provide a catalog of emission line identification
in the appendix.

\section{Model construction and calculation}\label{construction}

The model atmospheres consist of an underlying photosphere in radiative equilibrium, 
a chromosphere and part of the TR with a given temperature rise. 
All computations were carried out with the atmosphere code PHOENIX v13.7 \citep{phoenix}. 
For computations of M dwarfs chromosphere the proper NLTE treatment of the considered
species and the background opacities are especially important \citep{Andretta}, both of which are provided by
PHOENIX. Drawbacks in the chromospheric computations with PHOENIX are that no coronal flux
is -- yet -- incorporated and that all lines are computed with the assumption of complete redistribution.
Partial redistribution is especially important for the Ca H and K lines and for Ly$_{\alpha}$;
therefore none of these lines has been used in the modelling. Another assumption 
generally made in semi-empirical chromosphere modelling is hydrostatic  and 
ionisation equilibrium throughout the atmosphere. This assumption has been challenged recently
 for the Sun \citep[e.\,g.][]{Carlsson1}, but there are no detailed hydrodynamic chromospheric simulations for
M dwarfs available.

\subsection{The underlying photospheres}
For the
photospheres we used the best fit models to the red part of the UVES spectra which was
determined by a $\chi^{2}$ technique. The model grid for the comparison ranged from T$_{\rm eff}=2300 - 3700 $K
and log g $= 4.0 - 5.5$ in steps of $\Delta T_{\rm eff} = 100 $K and $\Delta$ log g$=0.5$. 
The photospheric log g and T$_{\rm eff}$ parameters determined in that fashion
are listed in Tab. \ref{phot}. We used solar chemical composition for all modelling  since
little is known even for the photospheric abundances in M dwarfs. In principle an abundance
analysis for the chromosphere is possible using the emission lines -- provided the temperature 
structure for the star is known. Such a technique would be especially useful for the late-type
M dwarfs where atomic photospheric lines are becoming rare. 
 
\subsection{Construction of the chromospheres}
For the chromospheres we used  the same ansatz as \citet{Short}, i.\,e., a linear
temperature rise vs. log\,$m$  with different gradients in the chromosphere and TR.
For some models we deviated from the linear temperature rise in
two different ways. For two stars we decided to alter the temperature
structure in the lower chromosphere to obtain a better fit of the Na\,{\sc i} D
lines (see Fig. \ref{temp}). 
This can be done without altering the flux in the other diagnostic lines
too much and is based on the gradient of models that fitted the Na\,{\sc i} D
line better than the model that gives the best overall fit.
 In the other case we tried to compose the chromosphere of two
linear segments, introducing a plateau in the upper chromosphere 
with the second linear part. This leads
to significantly more flux in the Balmer lines and these models were all
inferior to a single gradient temperature rise. However, the results of our
modelling suggests, that deviations from the linear temperature
rise may improve the fits, since different temperature gradients fitted
different lines best. For this paper we decided to stick to the linear
temperature rise, though  and obtain an overall best fit.
The top of the chromosphere is chosen at \mbox{8000 K} for all models since for
higher temperatures hydrogen is no efficient cooling agent any more \citep{Ayres}. 

\subsection{Construction of the TR}
The TR in our models
extends up to \mbox{$\log\,$T=5.0}. In principle the gradient in the TR should be given by 
Spitzer conduction. Since  conduction is balanced by the emerging flux from the TR,
the gradient can be computed if the flux from the TR is known. For late type M dwarfs there
are very few measurements for UV fluxes for temperatures below log T = 5.0. Therefore we
decided to model the TR temperature gradient as a function of the logarithmic column mass
($\log m$) semi-empirically as well, using the parameter
$\log \frac{dT}{d\log m}$. Balancing the conductive energy flux given by
$\kappa_{0} T^{5/2} \frac{dT}{ds}$ \citep[e.\,g.][]{Mihalas} with the emerging
radiation one obtains
\[\frac{d}{ds}\left(\kappa_{0} T^{5/2} \frac{dT}{ds}\right) = n^{2} P(T) \] with $P(T)$ denoting the cooling
function of a hot radiating plasma, $T$ the temperature and $\kappa_{0}\approx 
8.0\cdot 10^{-7}$ in cgs units \citep{Mihalas}. 
On neglecting the term involving second derivatives this leads to\[ \frac{dT}{d\log m} = m T^{-3/4} P(T)^{1/2} \cdot \rm const \]
Since $P(T) \sim T^{3/2}$ for 20\,000 $< T <$ 100\,000 and $m$ is about constant in the TR
our ansatz $\log \frac{dT}{d\log m}=\rm const$ is in good agreement with a conduction dominated
energy transport. The gradients we find in the best fit models also agree with the gradients
found by \citet{Jordan} for G/K dwarfs in the 50\,000 to 100\,000 K regime inferred from 
the emission measure distribution.

\subsection{Treatment of turbulent pressure}\label{turbulent}
The chromosphere models are computed 
in hydrostatic equilibrium on a given column mass grid including convection and
turbulent pressure.  
The treatment of the turbulent velocity $\xi$ and the resulting pressure  is
crucial for the models. We first used $\xi=2 \mathrm{km\,s^{-1}}$ in the photosphere
and a linear rise
to \mbox{10 $\mathrm{km\,s^{-1}}$} at the top of the chromosphere continuing into the whole
TR. However, the models with T$_{min}$ at higher pressure do not
converge for a linear turbulent velocity rise since it attains a significant
fraction of the sound velocity or even becomes larger than the sound velocity in individual
layers. Therefore the high pressure models are only possible for a description of
the turbulent velocity as a fraction of the sound velocity. We chose $\xi$ as
0.5 v$_{\rm sound}$. For the same model 
$\xi$ = 0.7 v$_{\rm sound}$ produces lower emission lines
than $\xi$ = 0.5 v$_{\rm sound}$ in the Balmer series and lower emission cores with self absorption
for the Na\,{\sc i} D lines. This is due to the lower electron pressure in the high turbulence velocity
model as discussed by \citet{Jevremovic}. A turbulent velocity of \mbox{2 $\mathrm{km\,s^{-1}}$} is applied to all photospheric
layers in all our models.  

\subsection{NLTE treatment}
Moreover we took into account NLTE effects for various species.
The models presented here normally treat H, He, \mbox{C {\sc i} - {\sc iii}}, \mbox{N {\sc i} - {\sc iii}}, 
\mbox{O {\sc i} - {\sc iii}}, \mbox{Fe {\sc i} - {\sc iv}},
Ti {\sc i} - Ti {\sc ii}, \mbox{Na {\sc i} - {\sc iv}} and \mbox{Mg {\sc i} - {\sc iii}} in NLTE. 
For some tests 
different NLTE sets of ions are chosen, which will be pointed out for these
models. All NLTE computations take into account all levels from the Kurucz database
\citep{Kurucz} or from the CHIANTI database \citep{Chianti} for C\,{\sc v} to C\,{\sc vi}, N\,{\sc vii} and
O\,{\sc v} to O\,{\sc viii} and Fe\,{\sc vi} to Fe\,{\sc vii}. 
Since NLTE computations show that the level populations of most species and especially
hydrogen are far from LTE we also took into account scattering for the LTE background lines,
since not all lines can be computed in NLTE. 
The scattering of the background lines is 
treated by choosing the $\varepsilon$ parameter in the relation S=$\varepsilon$B+(1-
$\varepsilon$)J, where S is the source function, B the Planck function and J the mean
intensity. If no scattering is included, i.\,e. $\varepsilon$=1.0, we obtain 
unrealisticly high emission in the Balmer lines for moderate chromospheres and for
stronger chromospheres even a Balmer jump and significant Balmer continuum, which is
not observed in any of our spectra. Since the NLTE level population strongly deviates from
the LTE level population in the chromosphere one would expect a lower value of $\varepsilon$.
Therefore we lowered $\varepsilon$ to 0.5 and to 0.1. The latter value gave no Balmer
jump and realistic Balmer line emission. Therefore we used $\varepsilon$=0.1 for all the
model computations. Since the $\varepsilon$ grid is rather coarse,
the uncertainties in this parameter are large. The model spectra needed clearly
$\varepsilon <$ 0.5 to avoid producing a Balmer jump, but for $\varepsilon <$ 0.5
it should be possible to balance the emergent emission with a TR onset at
lower column mass. For the best fit model of AD~Leo we computed also a model
with $\varepsilon$=0.2 which would lead to a lowering of the column mass at the onset of the
TR of about 0.1 dex.

\subsection{Filling factors}
All models are computed in spherical geometry. This means that we
assume an atmosphere with a filling factor of hot material of 100 percent. For the Sun this
assumption is under debate since the discovery of cool CO in chromospheric layers which lead
to thermally bifurcated atmosphere models \citep[e.\,g.][]{Comosphere}. For M dwarfs measured
magnetic filling factors are often very high: e.\,g. \citet{Saar} measured that about 73 percent
of the surface of AD~Leo is covered with active regions. For two other very active M4.5 stars
\citet{Johns-Krull} found magnetic filling factors of 50 $\pm$ 13 percent. 
We tried to infer chromospheric
filling factors a posteriori (see section \ref{filling}).

\subsection{Lines used for fitting}
To determine the quality of the 
fits different lines were chosen for the hotter and brighter stars AD~Leo, YZ~CMi and CN~Leo as compared to the
cool and very faint stars DX~Cnc and LHS~3003. For the three mid-type M dwarfs we used the 
Balmer line series higher than H$_{9}$, the Na\,{\sc i} D lines and a number of Fe\,{\sc i} and Mg\,{\sc i} lines 
between 3650 and 3870 \AA. An additional pronounced chromospheric feature is the He\,{\sc i} line at 5875 \AA, 
which could not be matched
for any of the three stars without creating too much flux in the Balmer lines. 
We ascribe this to our relatively simple approach with nearly linear temperature distributions
in the chromosphere and a linear temperature distribution in the TR. This problem applies not only
to the He line, but to other lines as well: it is hard to match large parts of the spectrum
well with the same model. Therefore our best fit models match most of the lines in principle
but not perfectly.

For the two late-type
M dwarfs we used the H$_{\alpha}$ and the H$_{\beta}$ line, the Na\,{\sc i} D lines and the 
He\,{\sc i} line at 5875 \AA\, since the blue part of the spectra were underexposed for the two
stars.


\section{The influence of the model parameters}

Before studying the five M dwarfs and their best fit model in detail a few words about
the influence of the general model parameters and assumptions are in order.

\subsection{Photospheric parameters}
The influence of the underlying photospheric model and its parameters T$_{\rm eff}$ and log g  
has been studied already by \citet{Short1}. They find that lowering or raising T$_{\rm eff}$
by 200 K or log g by 0.5 dex leads to uncertainties in the chromospheric parameters of
the colum mass at the onset of the TR or at the temperature minimum of about 0.3 dex. Therefore
we used individual photospheric models for each star. Nevertheless we  confirm the trend
they found comparing the models of AD~Leo and YZ~CMi (see section \ref{YZ}). 

Regarding the quality of our photospheric models we estimate the
error in T$_{\rm eff}$ to be about 100 K, since the variation of this parameter
led to signifacnt changes in $\chi^{2}$. A comparison of our T$_{\rm eff}$ 
values to those published in literature does indeed show agreement to
within 100 K (see Table
\ref{phot}). We also estimate the error in log g to be about 0.5 dex.

The influence of
the stellar mass of the underlying model has not yet been studied to our knowledge. We computed
models with 0.5 and 0.1 M$_{\odot}$ for our lowest mass star LHS 3003 and found only very weak dependence
on this parameter (see section \ref{LHS3003}). Also the influence of metalicity on the chromospheric emission has not
been studied so far. Fig. \ref{AD-Leo-3710} strongly suggests, that metalicity
should be included in future modelling if it uses metal lines as diagnostics. 

\subsection{Chromospheric parameters}
The microturbulence has been shown by \citet{Jevremovic} to influence
the line intensity as well as line shape. 
The influence of NLTE computations has been found in this article to have not only an
influence on emergent flux of the considered species itself, but on other species as well (see section 
\ref{NLTEsec}). 

The partial frequency redistribution approximation has not been implemented in PHOENIX so far.
\citet{Falchi} studied the impact of partial compared to complete frequency
redistribution (PRD/CRD) on the Ly$_{\alpha}$ line. They found that the PRD treatment of Ly$_{\alpha}$
also influences the Balmer lines. The CRD approximation leads to more emergent flux,
therefore our models should have the onset of the TR at too low a pressure compared to a model
computed in PRD. 

\begin{figure}
\begin{center}
\includegraphics[width=8cm]{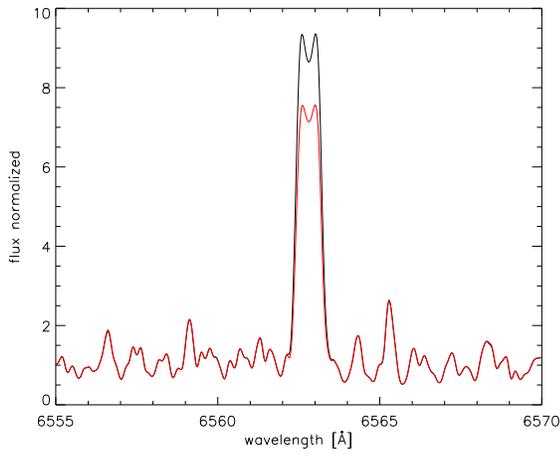}
\caption{\label{dust} Comparison of two models with dust (black) and without dust (red). The
continuum is normalized for both models, otherwise the difference in the H$_{\alpha}$
line would cease.}
\end{center}
\end{figure}

For the cool models with T$_{\rm eff}$= 2700 K and 2500 K dust may play an important role since
it warms the atmosphere in photospheric layers and is an efficient scatterer. For the
photospheric spectra themselves the influence of dust is not significant. For the chromospheres we
tried two different dust treatments usually used within PHOENIX. In the first approach the
dust is treated only in the equation of state, in the second approach the dust is also included
in the opacity calculations. While the first approach does not affect the emission lines, the
second approach enhances most of the emission lines  and lowers the continuum. A
comparison can be seen in Fig. \ref{dust}.
Although dust has an impact on the chromosphere modelling we did not use it 
in the present work since it slows down the computation significantly.  

As a last point the use of a linear temperature rise vs. log column mass in the chromosphere 
should be discussed. Alternatively one could use a non-linear temperature rise as 
was done by \citet{Mauas2000} or \citet{Vernazza} for the Sun using several diagnostics to
model different parts of the temperature rise. If there are enough diagnostic lines that
correspond directly to the temperature at a certain column mass this is undoubtedly  more
sensitive to the temperature structure and should give better agreement between data and
models. Since NLTE effects play an important role and
therefore certain lines can be affected by the radiation of layers far away from their
formation depth this is also a more complicated way to build chromospheres.
Therefore
we refrained from a non-linear temperature rise at the moment.

\section{Results}

\begin{figure} 
\begin{center}
\includegraphics[width=8cm]{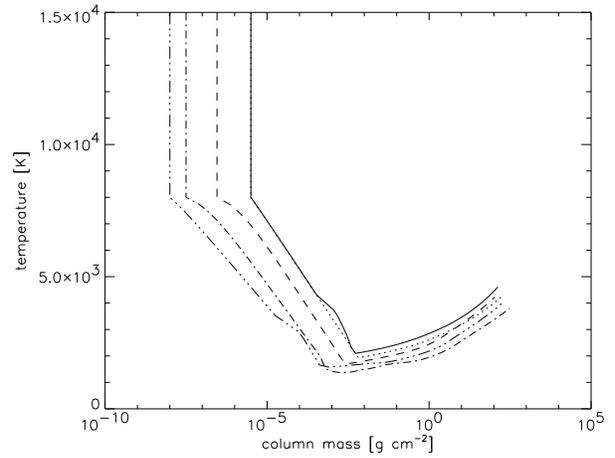}
\caption{\label{temp}Temperature structure for our best fit models. The solid line corresponds
to AD~Leo, the dotted line to YZ~CMi, the dashed line to CN~Leo, the dot-dashed line to LHS~3003
and the triple-dot-dashed line to DX~Cnc. The models for AD~Leo and YZ~CMi are nearly identical.}
\end{center}
\end{figure}

\begin{figure} 
\begin{center}
\includegraphics[width=8cm]{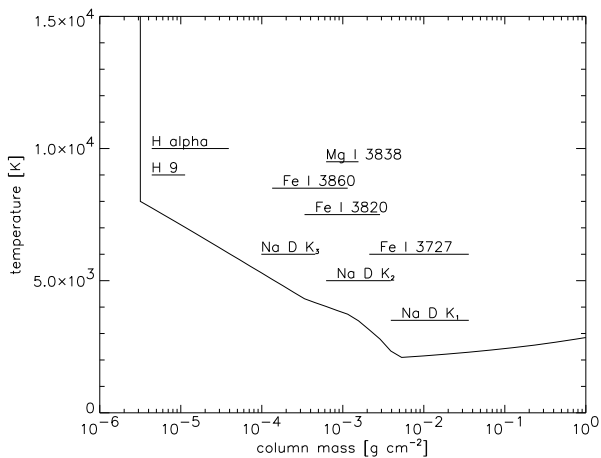}
\caption{\label{temp2}Temperature structure for our best fit models of AD~Leo. Given are the
approximate line formation depths for various lines used in the fitting process.}
\end{center}
\end{figure}

The temperature distribution for the best fit models is shown in Fig. \ref{temp} for all of
the stars and an approximate line formation depth is given for various lines for the
model of AD~Leo in Fig. \ref{temp2}. The pronounced emission line of He at 5875 \AA\,
is not indicated since the model of AD~Leo does not show this line. For the determination of the line formation depth we used
spectra for each layer in the atmospheric model. The formation of the resulting spectrum
in the outermost layer can be observed from layer to layer. One problem with this ansatz is
that the net flux is tracked, i.\,e. the flux going inward is accounted for as well. Therefore
the outer boundary of the line formation region can be determined quite well whereas the inner
boundary is less certain.

Since for all our observed spectra of the mid-type M dwarfs the Na {\sc i} D lines are found as absorption lines with
emission cores, the minimum seen in the line profiles can serve as a diagnostic for the
location of the temperature minimum in the star's atmosphere (as proposed by \citet{Andretta}
and \citet{Mauas2000}). We also find like \citeauthor{Andretta} that the Na {\sc i} D line profile
is insensitive to the TR gradient. For the three early-type M dwarfs the models show pronounced
self absorption in the emission core, which is not seen in the spectra, otherwise the Na {\sc i} D lines
are reproduced reasonably well for these stars (see e.\,g. Fig. \ref{YZ-Cmi-5880} for YZ~CMi).
For the two late-type M dwarfs the Na\,{\sc i} D lines are less well suited to diagnose the
temperature minimum since the absorption profile is quite shallow; this is especially true for
LHS~3003. 

The best fit was determined in two ways, by eye and with a $\chi^{2}$ test using
a number of wavelength ranges including the diagnostic lines. For LHS~3003 and DX~Cnc the
used wavelength ranges are 4855 to 4865, 6550 to 6570, 5873 to 5877 and 5888 to 5900 \AA.
For AD~Leo, YZ~Cmi and CN~Leo the $H_{\alpha}$ region was omitted and, instead, some of
the blue wavelength were used: 3705 to 3728, 3825 to 3840, 3780 to 3810 and 3850 to 3865. In the
case of AD~Leo and YZ~Cmi both the eye fit as well the $\chi^{2}$ fit resulted in 
the same best fit model. For CN~Leo the $\chi^{2}$ test
best fit model differed by 0.1 dex in the TR gradient to the one found by eye, i.\,e. they are
neighbouring models. Also for DX~Cnc and LHS~3003 the $\chi^{2}$ test preferred neighbouring
models to the by eye fit. We decided to use the models found by eye since for LHS~3003 the
$\chi^{2}$ test was contaminated by the Na D airglow lines and for DX~Cnc we preferred to describe
the $H_{\alpha}$ line more correctly than the $H_{\beta}$ line (which is the main difference between
the two models). 

We stopped improving the models when the variation of the three main parameters
(column mass at $T_{\rm min}$, column mass at onset of TR and grad TR) around some starting
parameters did not improve the fit. However, no true grid
in the parameter space was calculated, since we normally adjusted first the column mass at
$T_{\rm min}$ via the Na D lines, then varied the column mass at the onset of
the TR, readjusted the column mass at $T_{\rm min}$ and then varied
the gradient of the TR.   

\subsection{AD Leo} 

\begin{figure} 
\begin{center}
\includegraphics[width=8cm]{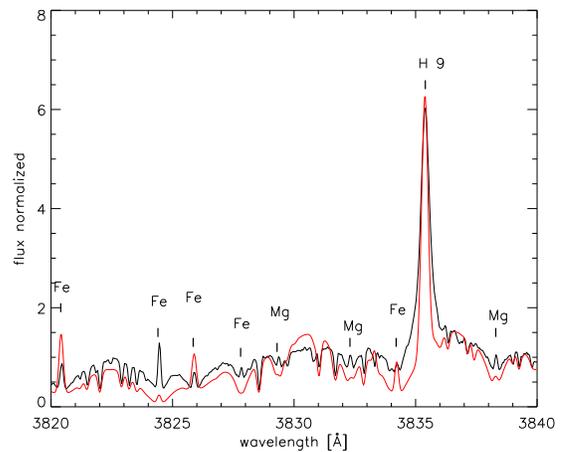}
\caption{\label{AD-Leo-3820}Comparison of the observed spectrum of AD~Leo (black) and the best 
fit model (grey/red). The emission cores/lines used for the modelling are indicated in the spectrum.}
\end{center}
\end{figure}

\begin{figure}
\begin{center}
\includegraphics[width=8cm]{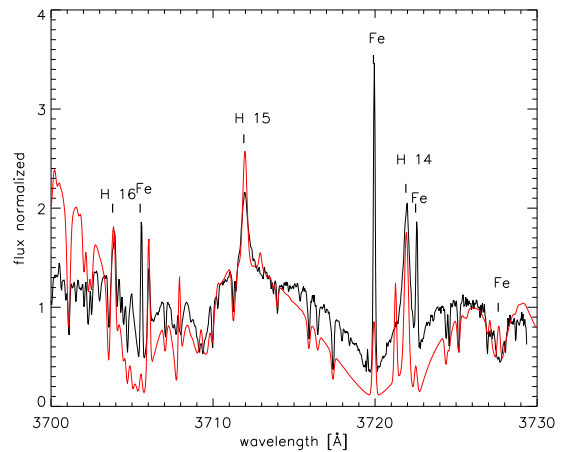}
\caption{\label{AD-Leo-3710}Same as is in Fig. \ref{AD-Leo-3820}. 
The absorption part of the Fe\,{\sc i} line at 3720 \AA, is too pronounced in the model what may be due to 
metalicity.}
\end{center}
\end{figure}

A comparison between model and data for the blue part of the spectrum of AD~Leo is shown in
Figs. \ref{AD-Leo-3820} and \ref{AD-Leo-3710}. The complex pattern of pure emission lines and
absorption lines with emission cores is reproduced quite well, although the amplitude of most of
the lines is not perfectly modelled. The base of the Balmer lines is somewhat broader than 
in the model but in general the Balmer lines fit
reasonably well in  amplitude and highest Balmer line seen. Iron lines can be either too strong
or too faint, but normally lines from the same multiplet behave in the same way. For example, the
three Fe\,{\sc i} lines at 3820, 3826 and 3834 \AA\, are all three from the same multiplet and too strong,
while the Fe\,{\sc i} lines at 3705.5, 3720 and 3722.5 \AA\, from another multiplet are all too faint.
 Since the deviation of the lines do not vary randomly it is unlikely that it
is caused by a lack of reliable atomic data; but even so one should keep in mind that especially
the collision rates are usually not well known. The behaviour may be caused
by our simple temperature structure. While the 3720 \AA\ line originates
in layers corresponding to temperatures between 4500 and 3500 K, the 3820 \AA\
line originates in layers corresponding to temperatures between 3800 and 2800 
K. The formation depths are overlapping with the 3820 \AA\, line forming in
deeper layers.
 An alteration of the temperature structure
in these layers should improve the fit of these lines. Nevertheless NLTE effects
also play  a role for these lines, since their amplitudes reacts  to
the gradient in the TR as well.

\subsection{YZ CMi}\label{YZ}

\begin{figure}
\begin{center}
\includegraphics[width=8cm]{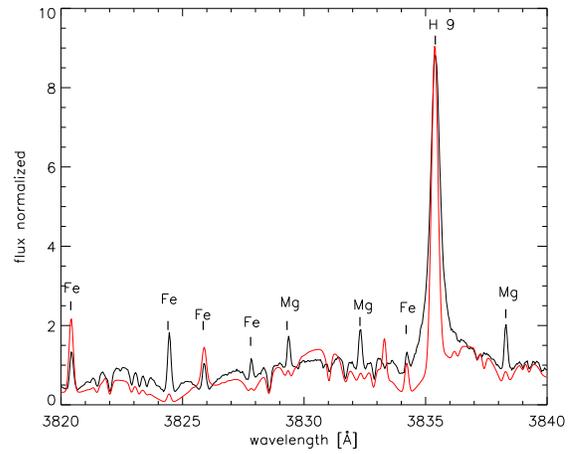}
\caption{\label{YZ-Cmi-3820}Comparison of the observed spectrum of YZ~CMi (black) and the best 
fit model (grey/red). The H$_{9}$ is reproduced quite well. 
The Mg\,{\sc i} lines show all less pronounced emission features in the model compared to the data.}
\end{center}
\end{figure}

\begin{figure}
\begin{center}
\includegraphics[width=8cm]{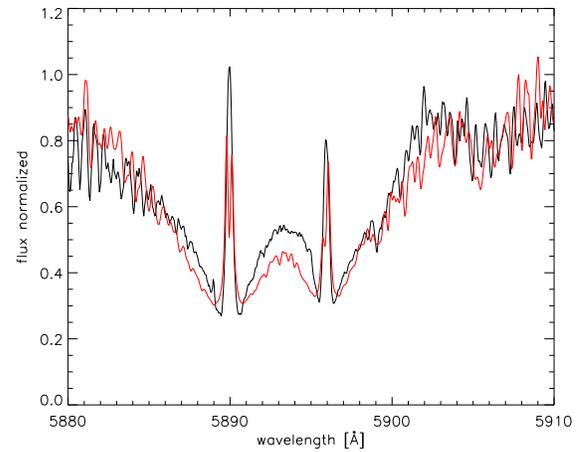}
\caption{\label{YZ-Cmi-5880}Same as is in Fig. \ref{YZ-Cmi-3820} but for the Na\,{\sc i} D lines. 
While the model shows deep self-absorption this is not seen in the data. }
\end{center}
\end{figure}

For YZ~CMi a comparison between model and data can be found in Figs. \ref{YZ-Cmi-3820} and
\ref{YZ-Cmi-5880}. The Balmer lines are well fitted  in amplitude. Also described
correctly is the highest Balmer line that is clearly seen in emission. The base width of the Balmer lines is too
narrow in the model. This is worse than for AD~Leo and comparable to CN~Leo. Since YZ~CMi is
relatively stable through the observations, pronounced activity is no obvious explanation for this. 
Since all three stars are very active this 
discrepancy may be caused by spiculae-like inflows and outflows (see also
sect. \ref{LHS3003}).

As AD~Leo, YZ~CMi shows a pattern of absorption lines with emission cores which is in general reproduced
well. The two stars are very similar, which is reflected in the similar models that differ only by
0.1 dex in the column mass at the temperature minimum and by 0.2 dex in the gradient of the
TR.  Judging from the spectra (observed or modelled), YZ~CMi is
the more active star since it shows more emission lines with larger equivalent width. 
Nevertheless, AD~Leo is described by the more active model judged by the temperature minimum 
located at higher column mass 
and the lower 
gradient in the TR.  However in combination with the higher effective temperature of the
photospheric model, it actually gives
lower amplitudes in the emission lines.
Even a temperature difference of 200 K in the effective temperature of the photosphere can
significantly influence the emergent chromospheric flux. Hence a good knowledge of the parameters
of the underlying photosphere is essential for chromospheric modelling.

Although the wings of the Na\,{\sc i} D lines are fitted reasonably well, the region between the
doublet lines is not. This region is  sensitive to the temperature minimum and the temperature
structure in the low chromosphere, but it is also sensitive to changes in the microturbulent
velocity. Also there are deep self-absorption cores in the Na\,{\sc i} D lines
seen in the models. This may be caused by a wrong temperature structure
in the mid chromosphere, where the flux in the line center of the Na \,{\sc i} D
lines arises (see Fig. \ref{temp2}), but the depth of the self-absorption core
in the Na\,{\sc i} D line is also strongly dependent on the NLTE set chosen.
It may therefore be a pure NLTE effect. In this case there is no strong connection
between the temperature structure of a certain part of the model and the
strength of the self-absorption. 

\subsection{CN Leo}

CN~Leo is the most active star in the sample and has accordingly the worst fit for the
individual lines. Nevertheless the disappearance of the absorption lines in favour of
pure emission lines is reproduced. While the only 100 K hotter photosphere of YZ~CMi
produces many absorption lines with emission cores, the spectrum of CN~Leo exhibits pure
emission lines - which is fully reproduced by the model. 

\begin{figure}
\begin{center}
\includegraphics[width=8cm]{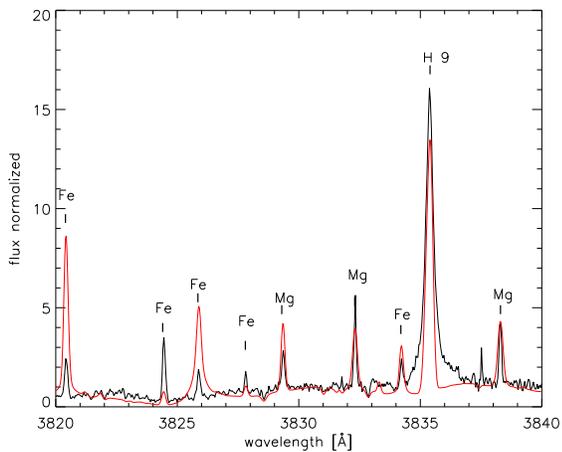}
\caption{\label{CN-Leo-3820} Comparison between data (black) and model (grey/red) for CN~Leo. Most of the Fe
lines are too strong in the model.}
\end{center}
\end{figure}

\begin{figure}
\begin{center}
\includegraphics[width=8cm]{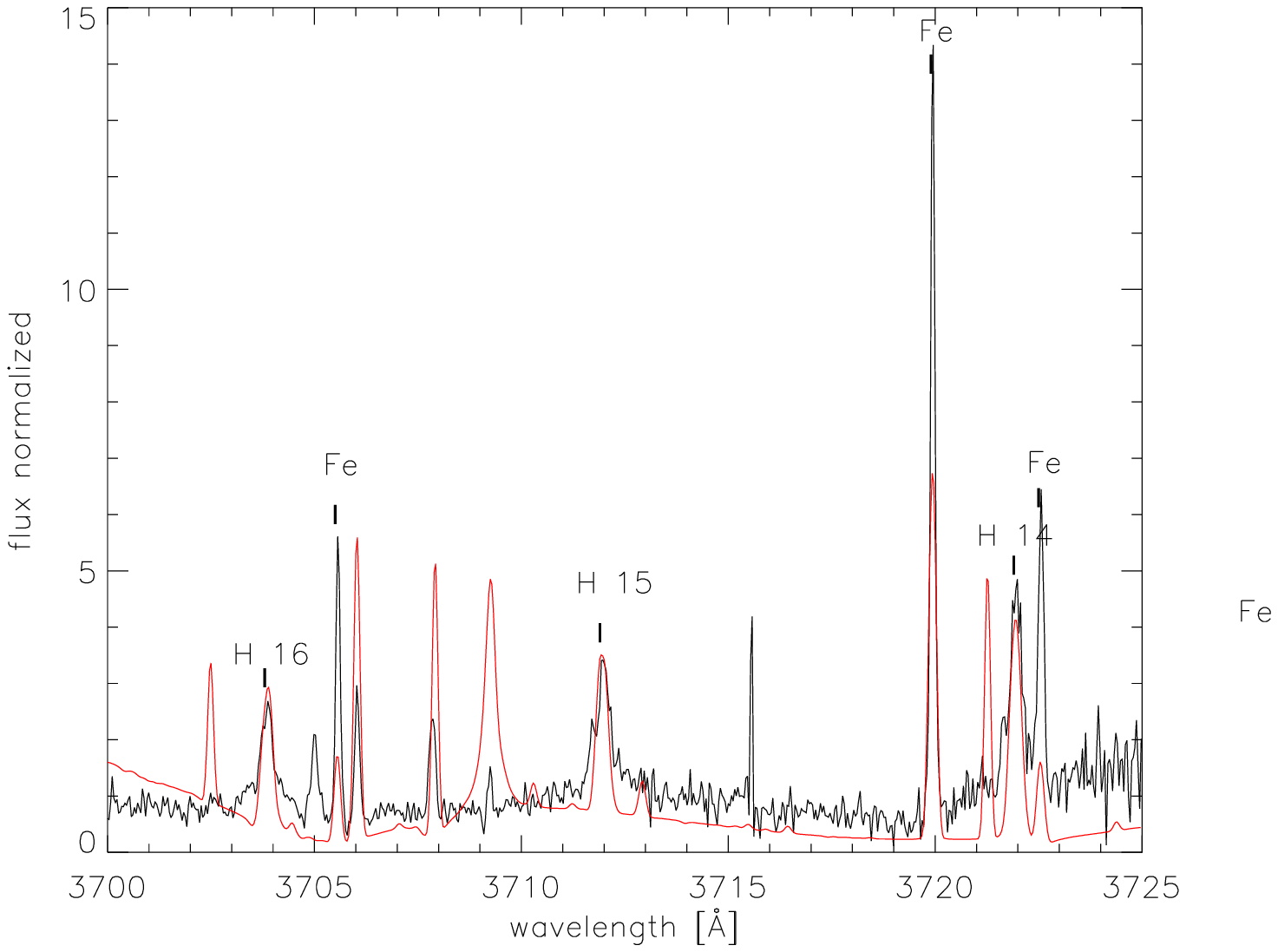}
\caption{\label{CN-Leo-3710}Same as is in Fig. \ref{CN-Leo-3820}. Again the Fe lines are too strong
while the three Balmer lines H$_{14}$ to H$_{16}$ are fitted quite well. }
\end{center}
\end{figure}

While the
amplitudes of the Balmer lines and the highest Balmer line seen are fitted quite well by the
model, the observed Balmer lines are much broader at the baseline than the ones in the model
(see Figs. \ref{CN-Leo-3820} and \ref{CN-Leo-3710}). Although the spectrum used for the modelling
is obtained during quiescence there may be some activity present  providing an
additional broadening mechanism. Moreover most of the Fe lines in the model are too strong compared
to the data if the Balmer lines are fitted well. On the other hand, models with well fitted iron lines yield
too much Balmer line flux. Fig. \ref{temp2} shows that the Fe lines are
forming in the middle and lower chromosphere, while the Balmer lines originate
from the top of the chromosphere. Therefore a non-linear temperature rise 
could help to solve this discrepancy. Again the model does not predict 
He emission lines. This incorrect modelling of the He\,D$_{3}$ line
may be caused by the non-inclusion of a corona in our models as suggested
by the work of \citet{HeSun}.

\subsection{DX Cnc}

\begin{figure}
\begin{center}
\includegraphics[width=8cm]{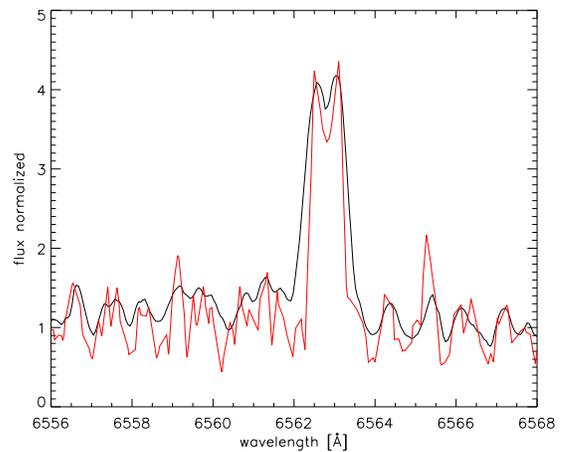}
\caption{\label{DX-Cnc-6550}Comparison of data of DX~Cnc (black) and best fit model (grey/red)
for the H$_{\alpha}$ line. 
The model shows too deep self-absorption. }
\end{center}
\end{figure}

While  the He\,{\sc i} line at 5875 \AA\, emission line is not predicted by the models for the three mid-type M dwarfs,
for the two late-type M dwarfs the line is too strong in the models. The emission cores of 
the Na\,{\sc i} D doublet are far too faint for this star using the linear temperature rise that
fit the Balmer lines well (see Fig. \ref{DX-Cnc-6550}). Therefore we modified the lower chromosphere temperature structure
and started with a higher gradient directly at the temperature minimum, but joined the original
temperature rise at about 4000 K before hydrogen starts to ionize. This leaves the Balmer lines
nearly unaltered and gives stronger emission cores in the Na D doublet.

\subsection{LHS 3003}\label{LHS3003}

While all other stars were computed with
0.5 M$_{\odot}$, LHS~3003 was computed with 0.1 M$_{\odot}$ because of its low effective
temperature and surface gravity and to test the influence of the mass on the chromospheric model.
While a higher mass produces a higher Ly~$\alpha$ line, the Balmer series lines are slightly lower. 
Therefore we conclude that the star's mass in the model has an influence on the chromospheric
model but the effect is small compared to other parameters and NLTE effects. 

\begin{figure}
\begin{center}
\includegraphics[width=8cm]{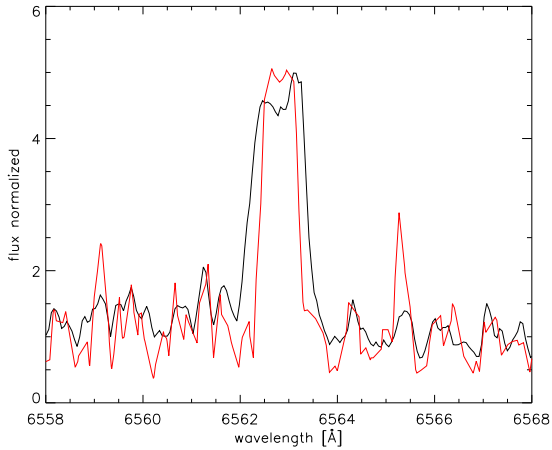}
\caption{\label{LHS-3003-6550}. Comparison of data of LHS~3003 (black) and the best fit model (grey/red) for the
H$_{\alpha}$ line. Here the self-absorption in the model is not deep enough. }
\end{center}
\end{figure}

\begin{figure}
\begin{center}
\includegraphics[width=8cm]{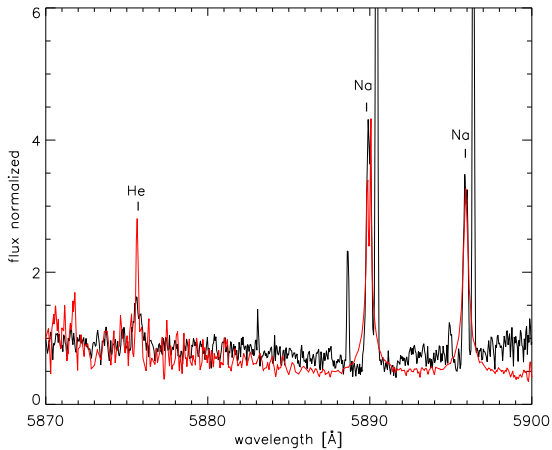}
\caption{\label{LHS-3003-5880}Same as is in Fig. \ref{LHS-3003-6550} but for the Na\,{\sc i} D lines
and the He\,{\sc i} line at 5875 \AA. The narrow emission lines are known airglow lines that have
not been removed (see also Fig. \ref{MASS}). }
\end{center}
\end{figure}

While the amplitude of the Na\,{\sc i} D lines is well fitted, 
the lines are too wide at the line base in the model (see Fig. \ref{LHS-3003-5880}). The width of the line
is also not fitted well for the H$_{\alpha}$, and for the H$_{\beta}$ line: These two lines are
too narrow in the model as can be seen in Fig. \ref{LHS-3003-6550} for the H$_{\alpha}$ line.
This may be due to the rotational velocity of the star. To test this hypothesis we spun up the
model to 20 - 30 km\,s$^{-1}$ to fit the line width of the
H$_{\alpha}$ line. On the other hand, we measured the
rotational velocity of LHS 3003 in the photospheric lines using CN~Leo as a template (see 
\citet{Fuhrmeister} for the method used); this procedure leads to a rotational velocity of 
\mbox{6.0$\pm$1.5
km\,s$^{-1}$}. If one compares the photospheric features in Fig. \ref{LHS-3003-6550} 
next to the H$_{\alpha}$ line to the model (with no rotation at all), a slow rotational
velocity seems to be very reasonable. Moreover this additional 
broadening is not seen in other emission lines e.\,g. the Na\,{\sc i} D lines.
Therefore rotational broadening can be ruled out and the chromospheric emission line must be affected by
another broadening mechanism. The modelling includes Stark and van der Waals broadening approximations. The self-broadening of the Balmer lines may, however, not be
described correctly \citep{Barklem}. 
Another possibility  ascribes the additional broadening to a more dynamic 
scenario with mass motions. If the star
hosts several active regions exhibiting mass motions, the overlapping of
the lines would lead to an additional broadening. Also spiculae-like
inflows and outflows may contribute to Balmer line broadening as described
for II~Peg by \citet{Pegasi}.
A dynamic scenario would also explain the asymmetric profile of both the H$_{\alpha}$
and the H$_{\beta}$ line in the star, which is not reproduced by any of our models. Since the 
H$_{\alpha}$ line shows two prominent components we tried to fit the line with two
Gaussians leading to Doppler shifts of -11 km\,s$^{-1}$ and 16 km\,s$^{-1}$, respectively, where
we used 6562.81 \AA\, as reference central wavelength. The H$_{\beta}$ line is not 
composed of two components, but fitting it with three Gaussian components leads to one component
at about rest wavelength and two components with Doppler shifts of about -11 km\,s$^{-1}$ and
18 km\,s$^{-1}$ using 4861.33 \AA\, as rest wavelength. Moreover the three averaged spectra show
some changes during the time series in the line profiles of the two Balmer lines. Since the
photometer shows no major flaring activity it seems that the quiescent activity of LHS~3003 
is composed of different active regions.

\section{Discussion}

\subsection{NLTE effects}\label{NLTEsec}

\begin{figure}
\begin{center}
\includegraphics[width=8cm]{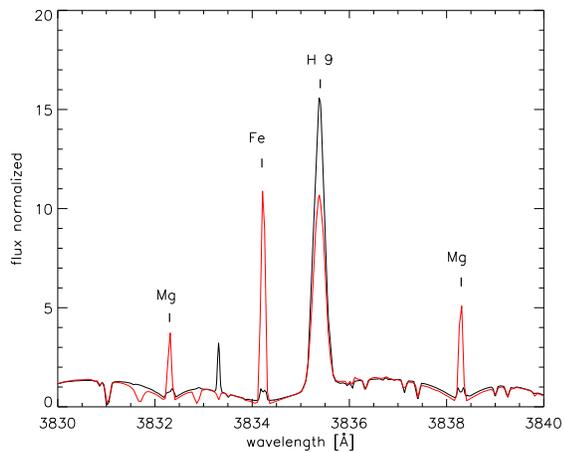}
\caption{\label{CNOcompH}Comparison of two computations with different NLTE sets. The line
at 3835.4 \AA\, is H$_{9}$, the other lines are Fe {\sc i} and Mg {\sc i}. The Fe {\sc i} and 
Mg {\sc i} lines in the grey/red model  often show much stronger
line flux than in the black model since they are not computed in NLTE in the grey/red model. 
The H$_{9}$ line
is influenced by the different NLTE sets of the two models, though hydrogen is computed in
NLTE for both models. Further computations show that this is caused mainly by the different treatment
of CNO in the two models.}
\end{center}
\end{figure}

The line flux and the line profile are both sensitive to NLTE effects. Therefore it
is necessary to compute at least all species in NLTE whose lines are used for the 
model fitting. 
Lines computed in LTE  usually show too strong emission as in Fig. \ref{CNOcompH}.
However, the lines are not only influenced by the NLTE calculations of the line forming species itself
but by other species as well. For example, the Balmer lines and the Na {\sc i} D lines
are influenced by the NLTE computations of carbon, nitrogen and oxygen (CNO). Therefore
as many species as possible should be treated in NLTE. 
To investigate the influence of different species on each other, 
we computed for YZ~CMi model atmospheres and spectra with different NLTE sets.
Hydrogen, helium and Na\,{\sc i} to Na\,{\sc iv} were always computed in NLTE. In addition we
computed the same model with CNO\,{\sc i} to CNO\,{\sc iii} in NLTE, and with C\,{\sc i} to 
\,{\sc vi}, N\,{\sc i} to N\,{\sc v}, O\,{\sc i} to \,{\sc vi} in NLTE. The latter NLTE set was
chosen to cover all ions of CNO with significant partial pressures in the temperature
domain below log T = 5.0. Moreover we computed the same model with different less abundant metals
in NLTE (in addition to CNO\,{\sc i} to CNO \,{\sc iii}): One set chosen was Fe\,{\sc i} to Fe\,{\sc vii}, Mg\,{\sc i} to Mg\,{\sc iv} and Ni\,{\sc i}
to Ni\,{\sc iii}, the other one Fe\,{\sc i} to Fe\,{\sc iv} and Co\,{\sc i} to Co\,{\sc iii}.
The latter two  models did not differ significantly in the Balmer lines nor in the Fe\,{\sc i}
lines. The difference in the Na\,{\sc i} D lines is less than about 10 percent. The electron
density for these two models are very similar except in a region around the temperature minimum,
which is seen in the variation of the Na\,{\sc i} D lines in the spectrum. The comparison of one
of these two models with CNO\,{\sc i} to CNO\,{\sc iii} in NLTE shows much larger differences both in
the electron density and in the spectrum. Therefore we conclude, that it is important to treat at 
least Fe\,{\sc i} to Fe\,{\sc iv} in NLTE. The largest changes can be seen in the amplitudes of
the higher Balmer
lines and in the Na\,{\sc i} D lines which is about 30 percent while for the H$_{\alpha}$ and
H$_{\beta}$ line the difference is about 10 percent. 

Even more important is the influence of the treatment of
the CNO ions. Between no CNO NLTE treatment at all and the first two ions in NLTE the amplitudes of
the Balmer lines change dramatically. While for the H$_{9}$ line the amplitude is about doubled, the 
amplitude of the H$_{\beta}$ line is about halved, the amplitude of the Na\,{\sc i} D lines is also
about halved. For further CNO ions treated in NLTE the flux in the  H$_{\alpha}$ and H$_{\beta}$
line decays further, while the amplitude of the Na\,{\sc i} D lines and the higher Balmer lines
stays nearly constant. The treatment of the higher CNO ions in NLTE is very problematic since
these ions are in principle not present in most parts of the atmosphere and therefore the code
must deal with very tiny numbers. 
 
The compensation for the additional flux in the models with the large NLTE sets 
were usually done via adapting the gradient in
the TR. Typically this gradient had to be increased for about 0.2 dex to compensate for the
additional emergent flux in the Balmer series. 

\begin{figure}
\begin{center}
\includegraphics[width=8cm]{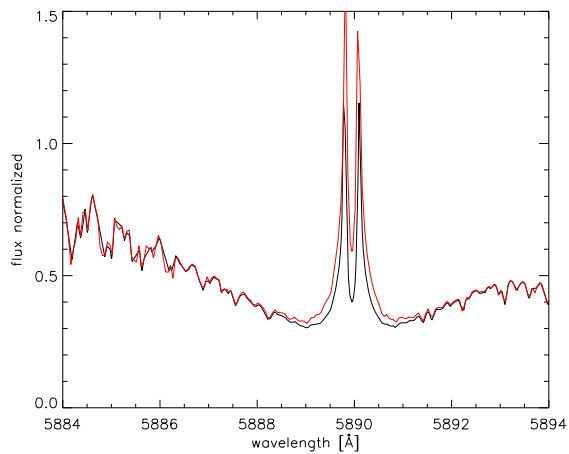}
\caption{\label{CNOcompNa}Same comparison as in Fig. \ref{CNOcompH} for the Na\,{\sc i} D line.
The NLTE set that also includes Fe, Mg and CNO show shallower Na emission (black line) than the
one that treats only H, He and Na in NLTE (grey/red line). 
}
\end{center}
\end{figure}

\begin{figure}
\begin{center}
\includegraphics[width=8cm]{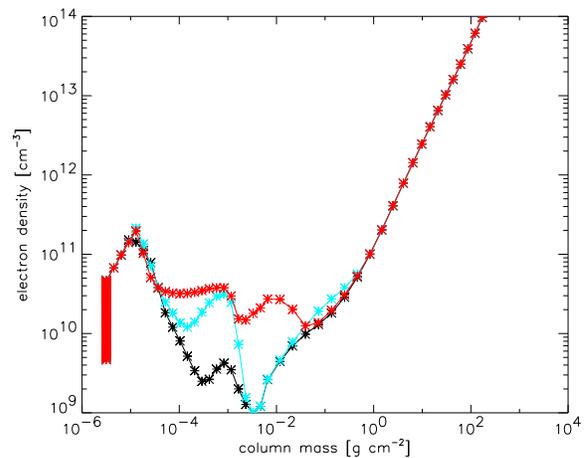}
\caption{\label{necomp} Comparison of the influence of different NLTE sets on the electron
density. The black line is the basic model with H,\,He and Na in NLTE. Grey/red is the model 
with Fe/Ni in NLTE and light grey/turquoise is the model with CNO\,{\sc i} to CNO \,{\sc iii}
in NLTE. All three models differ significantly in large parts of the atmosphere. }
\end{center}
\end{figure}

One explanation of this effect is through
the direct influence of the NLTE calculations on the electron density: NLTE calculations of
one species influence the electron density and therefore the level population of all other
species. Another possibility is the influence of a species with a multitude of lines
in the hydrogen continuum region which would influence the hydrogen ionisation equilibrium
and therefore indirectly the electron density. Since cobalt and nickel  have many lines
in the hydrogen continuum region and these species do not influence the electron density
too much, this seems to be a second order effect compared to the direct influence of the
electron density via the ionisation balance of the more abundant elements (see Fig. \ref{necomp}).  
On the other hand, the NLTE computation of the Fe ion levels influence the electron density in
large parts of the atmosphere. 

To gain some further insight in the NLTE behaviour of chromospheres we built a simple chromosphere model for the Sun
that is supposed to be closer to LTE than M dwarfs. We computed a model with H, He and Ca\,{\sc i}
- {\sc v} and a model with H, He, Ca\,{\sc i} - {\sc iii}, CNO\,{\sc i} - {\sc iii}, and 
Fe\,{\sc i} - {\sc iii} in NLTE. The two models show very little differences in the Balmer lines
and in the Ca\,{\sc ii} H and K line. The largest variation seen in these lines is about 10 percent
 which is indeed much less than in the M dwarfs. Therefore the strong
 variations due to the set of NLTE species chosen is characteristic of
 the M dwarf models. Although this does not rule out
computational artefacts it strengthens our confidence in the reliability
of the variations in the electron pressure and the spectral lines
in our M dwarfs.

\subsection{Uniqueness of the models}
During our chromospheric modelling we encountered the problem of non-uniqueness of the models in
several ways, but it could be resolved each time. Since the Na D lines are known to be only sensitive
to changes in the lower chromosphere, clearly an additional diagnostic is needed. The Balmer
lines are sensitive to changes in the upper chromosphere and the gradient in the TR. Nevertheless
we found models with nearly identical Balmer lines either in the lines higher than H$_{9}$ or in
H$_{\beta}$ and H${_\alpha}$ but normally not both. Therefore the very high Balmer lines seems to
be an important diagnostic tool. Only for different NLTE sets the whole Balmer series
can be nearly identical for different model parameters. Since e.\,g. the Fe lines are
not reproduced correctly by a LTE treatment, the models can be distinguished by these lines.

Since we used only H$_{\alpha}$, H$_{\beta}$ and the Na D lines for the model construction of
  LHS~3003 and DX~Cnc, these models may be affected by non-uniqueness. However since the emission
lines react more sensitively to all chromospheric parameters when lowering T$_{\rm eff}$
we do not regard this as a serious problem.

Another interesting point is to test if the photospheric parameters used may
introduce some kind of non-uniqueness. Since AD~Leo and YZ~CMi have the most
similar photospheric spectra in the sample, we tested if the  spectrum
of YZ~CMi could be fitted with one of the model spectra created for AD~Leo
that has a hotter underlying photosphere than YZ~CMi and the same log g.
We found that the models with the hotter photosphere give poorer fits since
a stronger chromosphere was needed to match the Balmer lines, which led to
too much flux in the metal lines. Since the chromospheric parameter space is
designed for AD~Leo, though, we may miss a well fitting model with a
hotter photosphere for YZ~Cmi. 

\subsection{Comparison to other models}

Our hottest sample star AD~Leo is the only one in our sample for which  chromosphere models have been
computed
by other authors. \citet{MauasADLeo} derived a semi-empirical model of the photosphere and 
chromosphere of the
quiescent state of AD~Leo fitting the continuum as well as some chromospheric lines. 
\citet{Hawley} constructed a grid of flare models for AD~Leo including a quiescent one. These
models were constructed on a photospheric base by using X-ray heating from a model for the
overlying corona. \citet{Short1} constructed chromospheric and TR models in 
a very similar way as our approach, using a linear temperature vs. log $m$ distribution. Moreover, all these authors
used very different underlying photospheres. While \citet{MauasADLeo} fitted both photosphere 
and chromosphere at the same time semi-empirically and no T$_{\rm eff}$ or log g values were given, \citeauthor{Hawley}
used a model by \citet{Mould} with T$_{\rm eff}=3500$ K and log g = 4.75, and \citeauthor{Short1}
used a PHOENIX photosphere with T$_{\rm eff}=3700$ K and log g = 4.7. All authors used solar chemical
composition as we did. Since the emission lines are sensitive to the underlying photospheres as mentioned
above and discussed in \citet{Short1}, and due to the different construction all these models are 
difficult to compare. The models of \citet{Short1} are very similar to our own models, 
although they used
a photosphere considerably hotter. Accordingly they found a TR onset at higher pressure than
we did. 

\subsection{Filling factors}\label{filling}
Our one dimensional model atmospheres correspond to a 100 percent filling factor of chromospheric
material. Nevertheless one can estimate the filling factor a posteriori. With distance and radius
of the star known, the theoretical scaling factor between flux at the stars surface and flux at
earth atmosphere can be calculated. This can be compared to the scaling factor between the model
and the flux calibrated spectrum; the ratio of the two gives the filling factor of the
chromosphere. We calculated the radius of the stars from the measured luminosity and the temperature
of the photospheric model via $L_{\rm bol}=4\pi R^{2}\sigma T_{\rm eff}$. The largest uncertainty is
caused by the flux calibrations since the standard star was observed only once per night. We
estimate the error in the absolute fluxes to be as large as a factor of two. The calculated
filling factors can be found in Table \ref{filltab}. Due to the possibly big errors in flux 
calibration the filling factors should be considered only as estimates, nevertheless, the three
early and mid-type M dwarfs seem to have higher filling factors than the two late-type dwarfs.

The low chromospheric filling factors of the two late-type M dwarfs
imply that  one-dimensional model 
calculations may be a poor approximation, since a significant portion of
the energy in the active regions will be radiated away horizontally. 
Therefore  stronger
chromospheres than we have inferred here may be embedded in cool material.
For proper simulations of such a chromosphere three-dimensional hydrodynamical
simulations would be needed, as is done for the Sun, e.\,g. by \citet{Wedemeyer}. 

\begin{table*}
\caption{\label{filltab}Filling factors of the chromospheres for the modelled stars and the
parameters used to calculate the filling factors. The measured scaling factor $f_{m}$ between
model and flux calibrated spectrum is given as well as the theoretical scaling factor
$f_{th}= R^{2}/d^{2}$.}
\begin{tabular}[htbp]{ccccccc}
\hline
star & $\log L_{bol}$& radius [cm] & distance [pc] & $f_{m}$ & $f_{th}$ & filling factor\\
\hline
AD Leo & 31.94 $^{1}$& 3.4  $10^{10}$& 4.9 $^{1}$& 3.0 $10^{-26}$ & 5.3 $10^{-26}$ & 0.57\\
YZ CMi & 31.69 $^{1}$& 2.9  $10^{10}$& 6.1 $^{1}$& 1.7 $10^{-26}$& 2.5 $10^{-26}$ & 0.68\\
CN Leo & 30.55 $^{1}$& 8.4  $10^{9}$& 2.4 $^{1}$& 0.7 $10^{-26}$& 1.4 $10^{-26}$ & 0.50\\
DX Cnc & 30.48 $^{1}$& 8.9  $10^{9}$& 3.6 $^{1}$& 1.7 $10^{-27}$ & 6.6 $10^{-27}$ & 0.26\\
LHS 3003 & 30.31 $^{2}$& 8.5 $10^{10}$ & 6.4 $^{3}$& 0.5 $10^{-27}$ & 1.9 $10^{-27}$& 0.26\\
\hline
\end{tabular}
\\
$^{1}$ \citet{Delfosse} \hspace*{1cm}
$^{2}$ \citet{Leggett} \hspace*{1cm}
$^{3}$ \citet{Dahn}
\end{table*}


\section{Summary and conclusion}

We have presented the first semi-empirical models for mid- and late-type M dwarfs accounting
for Fe\,{\sc i} lines. We found that the models are able to fit the transition from absorption
lines with emission cores to pure emission lines at about 3000/2900 K effective temperature.
Moreover we found models for five individual stars with effective temperature between 3200 and
2500 K that fit many spectral features reasonably well. We found for Fe\,{\sc i} lines from 
the same multiplet that they normally behave the same way: The model
predicts too high or too shallow amplitudes for all of them. Therefore some of the multiplets
may be used to obtain more complex temperature distributions that are able to fit the spectral
lines even better. 

Moreover we investigated the behaviour of the emerging spectrum under NLTE calculations of
different species and found a large influence especially of CNO, that may alter the Balmer
and Na\,{\sc i} D lines significantly. There is far less NLTE crosstalk caused by species
 like Co, Ni and Ti.

For the two late-type M dwarfs DX~Cnc and LHS~3003 we studied the behaviour of the models
if dust is considered and found that dust can affect the emerging emission lines if it is
considered not only in the equation of state but also in the opacity calculations.

These very late type objects can be described by the same type of
chromospheric model atmosphere as the earlier M dwarfs except that the onsets of
the chromosphere as well as the TR move to lower pressure. 
Thus, there seem to be no principal differences in the heating mechanisms of the
chromospheres down to M7. This is relevant in the context of the ongoing discussion
about the decreasing activity of the very late-type objects 
and in particular whether they
have possibly only transient chromospheres and coronae. Since the decrease in activity starts at
around M7, even more late-type objects than hitherto investigated should be included in
chromospheric modelling attempts.

We therefore conclude that modelling of chromospheres with semi-empirical deduced
temperature distributions relies heavily on correct input parameters and model assumptions
such as effective temperature of the photosphere, log g, NLTE treatment of important lines
and dust treatment for the coolest stars. Another parameter probably as important as the others
is the metallicity which is normally not considered in chromospheric modelling.


\begin{acknowledgements}

Most of the model computations were performed at the Norddeutscher Verbund
f\"ur Hoch- und H\"ochstleistungsrechnen (HLRN) and at the Hamburger Sternwarte
Apple G5 cluster financially supported by a HBFG grant. \\B.~Fuhrmeister acknowledges
financial support by the Deutsche Forschungsgemeinschaft under DFG SCHM 1032/16-3,
and thanks A. Schweitzer for numerous discussions that contributed to this work. 
PHH was supported in part by the P\^ole Scientifique de Mod\'elisation
Num\'erique at ENS-Lyon.  

\end{acknowledgements}


\begin{appendix}
\section{Identification of chromospheric emission lines}\label{appendix}

Chromospheres of M dwarfs exhibit hundreds of emission lines especially in the near
UV between 3000 and 4000 \AA. These lines can even be multiplied during flares. Deciding
which lines to use for the chromospheric modelling made us produce an extensive emission 
line list for the 
near UV and optical. We restricted that list not to the 5 modelled M dwarfs  but 
used the
whole sample of 21 late-type stars and brown dwarfs. 506 different emission
lines of the elements H,\,He,\,Na,\,Mg,\,Al,\,Si,\,K,\,Ca,\,Sc,\,Ti,\,Cr,\,Mn,\,Fe,\,Co,\,Ni 
could be identified,
revealing the different levels of activity in the stars. 

The observation parameters can be
found in Tab. \ref{starlist}. For further information about the observations and the
data analysis see the article above or \citet{Fuhrmeister}.

\begin{table*}[!ht]
\caption{\label{starlist}Basic observations parameters of the observed stars.}
\begin{tabular}{lllll}
\hline
name & other & spectral  &  observations &number of\\
     &  name & type &                  &identified lines  \\
\hline
LHS 1827& GJ 229A&M1 & 2002-03-15 4 spectra 1200s&11\\
LHS 5167 & AD~Leo & M3.5 & 2002-03-13 3 spectra 1800s& 142\\
         &        &    &2002-03-16 2 spectra 1200s& \\
HD 196982 & AT~Mic & M4.5 & 2002-03-16 2 spectra 2400s& 91\\
LHS 1943 & YZ~CMi & M4.5e& 2002-03-13 3 spectra 3600s& 178\\
LHS 2664 & FN~Vir & M4.5 & 2002-03-13 3 spectra 3600s& 74\\
LHS 324 & GL~Vir & M5 & 2002-03-13 3 spectra 3600s& 72\\
        &        &    &  2002-03-16 2 spectra 2400s& \\
LHS 36 & CN~Leo & M5.5 & 2002-03-13 6 spectra 7200s& 244\\
       &        &      & 2002-03-14 4 spectra 4800s& \\
       &        &      & 2002-03-15 6 spectra 7200s& \\
       &        &      & 2002-03-16 6 spectra 7200s& \\
       &        &      & 2001-01-06 1 spectrum 3120s& \\
LHS 2076 &EI~Cnc & M5.5  & 2002-03-15 4 spectra 4800s& 80\\
         & &      & 2002-03-16 1 spectrum 1200s& \\
LHS 49 & Prox~Cen & M5.5 & 2001-02-02 1 spectrum 3120s& 147 \\ 
LHS 10 & UV~Cet & M5.5 &2000-12-17 1 spectrum 3120s& 109\\
LHS 248 & DX~Cnc &  M6 & 2002-03-16 3 spectra 3600s& 17\\
LHS 2034 &AZ~Cnc & M6 & 2002-03-14 6 spectra 6000s& 251\\
         & &    & 2002-03-16 2 spectra 2400s& \\
LHS 292 &  & M6.5& 2001-02-02 1 spectrum 3120s& 103\\
LHS 429 & vB~8 & M7 & 2002-03-13 3 spectra 3600s& 23\\
        &      &    & 2002-03-15 3 spectra 3600s& \\
LHS 3003 & & M7 & 2002-03-14 3 spectra 3600s& 7\\
LHS 2397a & & M8 &  2002-03-14 3 spectra 3600s& 9\\
LHS 2065 & & M9 & 2002-03-13 3 spectra 3600s& 7\\
DENIS-P J104814.7-395606& &M9 &2002-03-14 4 spectra 4800s& 4\\
DENIS-P J1058.7-1548& & L3 &2002-03-15 4 spectra 4800s& -\\
2MASSI J1315309-264951& & L3 &2002-03-15 3 spectra 3600s& 4\\
Kelu-1& CE 298& L3 & 2002-03-14 3 spectra 3600s& 2\\
      &       &  &2002-03-16 3 spectra 3600s& \\
\hline
\end{tabular}
\end{table*}

IRAF was used to measure the central wavelength, FWHM and the equivalent
width (EW) of the emission lines. We decided to fit the background by eye since otherwise
emission cores could not be treated and there are wavelength ranges where the lines are
so crowded that it is hard to find appropriate pseudo-continuum points. Therefore the EW measurements
are affected by a rather large error. For single lines in spectral wavelength ranges with an ill
defined continuum this error may be as big as a factor of two, but for most lines it is less
than 40 percent.  

We identified the emission lines with the help of the Moore catalog \citep{Moore}. Few lines
were identified via the NIST database \footnote{available online under\\ 
http://physics.nist.gov/cgi-bin/AtData/main\_asd}. Moreover
we re-identified a random sample of about 5 percent of the lines with the help of the PHOENIX atmosphere models
and found full agreement. 
Of the identified lines
358 are in the blue arm and 148 are in the red arm of the spectra. In the electronically
published Tab. A.2 the central wavelength of the lines as well as the FWHM and the EW 
can be found. 

All spectra besides the brown dwarfs were shifted to rest wavelength
before the measurements of the lines. Therefore the main problem was the recognition what is
an emission line or emission core. This suffers from the low signal to noise ratio especially in the blue end
of the spectra where the count rates are very low. In addition  in some wavelength ranges
the continuum is not well defined and in these regions it is sometimes hard to decide
whether a particular feature is an emission line or barely left over continuum between
consecutive absorption lines. 
Since we excluded doubtful features from our line list, the list
 cannot be claimed to be complete for weaker lines.

We provide some remarks on individual stars:

\subsection{GJ 229A}
This is the earliest star in the sample and rather inactive. It shows only Fe {\sc i} and
Fe {\sc ii} emission cores and no pure emission lines at all.

\subsection{AD~Leo, YZ~Cmi, Prox~Cen and LHS 292}
The activity level from AD~Leo can be compared to YZ~CMi, Prox~Cen and LHS~292. 
All three stars exhibit
about the same number of Fe {\sc i}, Fe {\sc ii} and other metal lines, but AD~Leo and LHS~292
show very few
Ni~{\sc i} emission lines compared to the other two stars. LHS~292 also shows a reduced number
of Ti\,{\sc ii} lines.

\subsection{AT~Mic, FN~Vir and GL~Vir}
The reduced number of emission lines in these stars is at least partly due to the much
lower S/N ratio compared to AD~Leo, YZ~Cmi and Prox~Cen.

\subsection{UV~Cet}
UV~Cet is a fast rotator with v\,sin i=32.5 $\mathrm{km\,s^{-1}}$ \citep{Mohanty}. Superimposed
on the broad emission lines is a set of slightly redshifted, narrow emission lines. We
tentatively interprete this as emission from a chromospheric active region \citep{Fuhrmeister}.

\subsection{CN Leo}
Besides the line list for the averaged spectrum of CN Leo we produced line-lists for a one hour
averaged time series on 2002 March, 16th containing a short duration flare and a line list for an one hour
averaged time series on 2002 March, 15th showing enhanced emission line flux but no pronounced variablity
in the light curve. Since the flare was only a short duration flare this line list is showing actually
less emission lines than the averaged one, since CN~Leo showed less emission lines directly before and
after the flare.

\subsection{LHS~2076}
The star exhibited a short duration flare during the observations and for the single flare spectrum
an additional line list was created. While the number
of the Fe\,{\sc i} lines remain nearly unchanged during the flare, the number of Ti\,{\sc ii} lines increases
from 2 to 11 and the number of Ni\,{\sc i} from 4 to 13 lines revealing temperature
changes in the chromosphere during the flare. During the flare the lines are statistically
 redshifted by an amount of about 0.05 \AA.

\subsection{LHS~2034}
The star underwent a long duration flare during the observation. We created an additional line list
for the first spectrum taken containing the flare peak. This flare spectrum shows
particularly many emission lines in the red part of the spectrum -- where the other stars show
only very few lines. No line shifts could be found for the flare spectrum. 
Since the flare decays very slowly the averaged spectrum is affected by the
flare spectrum. 

\subsection{LHS 3003, LHS 2397a, LHS 2065, vB-8, DX~Cnc and DENIS-P J104814.7-395606}
Though the blue part of the spectrum is underexposed for these stars we could identify some
emission lines. Moreover DENIS-P J104814.7-395606 showed flaring activity during the observations which led to blueshifts
in the spectral lines. This star is discussed in detail by \citet{Fuhrmeister1}. 

\subsection{DENIS-P J1058.7-1548, 2MASSI J1315309-264951 and Kelu-1}
No radial velocity correction was applied to these very late-type objects, since we had no
appropriate template spectrum. Therefore the central wavelength given for these objects is
not corrected for radial velocity. 

\begin{figure}
\begin{center}
\includegraphics[width=8cm,height=5cm,clip=0]{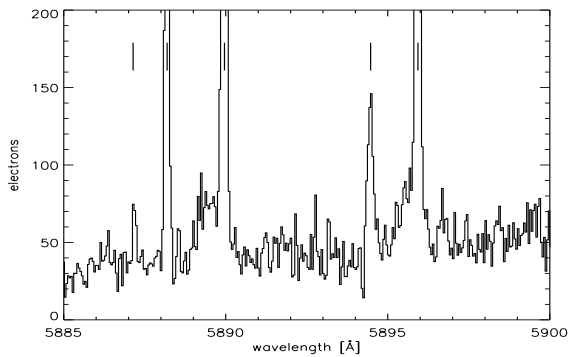}
\caption{\label{MASS} Spectrum of 2MASSI J1315309-264951 around the Na\,{\sc i} D lines. Known
airglow lines for the UVES instrument are marked with a vertical line.}
\end{center}
\end{figure}

For the L3 dwarf 2MASSI J1315309-264951 we report besides the detection of 
H$_{\alpha}$ and H$_{\beta}$ weak Na\,{\sc i} D emission lines (see Fig. \ref{MASS}). 
This is to our knowledge 
the first detection of the Na\,{\sc i} D lines in emission in an L dwarf. 
These are heavily blended with airglow lines known for the UVES instrument \citep{Hanuschi}.
The EW for H$_{\alpha}$ of 24.1 \AA\, seems to imply that the brown dwarf is in a rather
quiescent state since \citet{Hall} found EW of 121 \AA\, and of 25 \AA\, half a year later.

\end{appendix}

\bibliographystyle{aa}
\bibliography{papers}

\end{document}